\documentclass[12pt,a4paper]{article}
\usepackage[dvips]{epsfig}
\usepackage{afterpage}

\begin{document}    %--------------------------------------
\begin{rmfamily}

\begin{center}
{\Large  {\bf $\sigma(600)$ and background in $\pi\pi$ scattering} }\\[0.5cm]
{\bf  A.E.Kaloshin \footnote{EM:\ kaloshin@physdep.isu.ru},
V.M.Persikov and A.N.Vall  } \\
{\it  Irkutsk State University,  K.Marks Str., Irkutsk 664003,
Russia
}\\[0.5cm]
\end{center}

\begin{abstract}
We suggest a simple analytical description of the S-wave isoscalar
$\pi\pi$ amplitude, which corresponds to a joint dressing of the
bare resonance and background contributions. The amplitude
describes well the experimental data on the $\delta^0_0$ phase
shift in the energy region below 900 MeV and has two poles in the
$Re\ s >\ 0$ half-plane. Besides the well-known pole of
$\sigma(600)$-meson with $Re\ s \sim m_{\pi}^2$, there exists a
more distant pole with $Re\ s \sim 0.6 \ GeV^2$. Our analysis
indicates for the dynamical origin of the $\sigma(600)$ pole,
while the second pole should be associated with lowest $q\bar{q}$
state.
\end{abstract}

\section{Introduction}

The properties of the lightest scalar meson $\sigma(600)$ are very
important for interpretation of a scalar family and details of the
chiral symmetry breaking. Appearance of a new experimental
information  and the theory development in the low energy region
generated an extensive discussion on this issue (see
\cite{Min,Ani00,Kle,Tor,Bev} and references therein). As for
existence of $\sigma(600)$, now it is a commonly accepted fact and
this resonance is included back into Particle Data Group's tables.

There is a long story concerning the resonance interpretation of
the S-wave amplitude $\pi\pi\to\pi\pi$ with isospin I=0. One of
the key moments of this story was a realization (see e.g.
\cite{Ach94, Ish, Tak}) that apart of the resonance $\sigma(600)$
term there is an essential background contribution in this energy
region. However there is no evident recipe to divide the $\pi\pi$
amplitude into the resonance and background terms. The simplest
and widely used method is "adding in phase shift" of the resonance
an background contributions (IA method in terminology of
\cite{Ish, Tak}):
\begin{equation}
\delta^0_0 = \delta^R + \delta^B .
\label{add}
\end{equation}
The amplitude $\pi\pi\to\pi\pi$ in this case is:
\begin{eqnarray}
f^0_0(s)&=&\frac{e^{2i\delta^0_0}-1}{2i\rho}=
\frac{e^{2i\delta^B}-1}{2i\rho} +
e^{2i\delta^B} \cdot \frac{e^{2i\delta^R}-1}{2i\rho}=
\nonumber \\
&=&\frac{e^{2i\delta^B}-1}{2i\rho} + e^{2i\delta^B}\cdot f^{Res}(s) .
\label{IA}
\end{eqnarray}

The anzats (\ref{add}) may be derived from summation of the loop
contributions with some extra conditions \cite{Ach94}. To obtain
the resonance parameters from the experimental data one needs an
additional assumption about the form of the background
contribution $\delta^B$. The best way for a broad resonance is to
determine its mass and the width from the pole position in the
complex energy plane. However only the resonance contribution
$f^{Res}(s)$ of the entire amplitude (\ref{IA}) can be continued
into the complex energy plane. Thus it is possibly to study the
pole position but not the pole residue.

Other methods to describe the $\delta^0_0$ phase shift, different
from (\ref{add}), either have the so evident defects, or are much
more complicated with many free parameters.

In this paper we suggest a very simple analytical parameterization
for $\pi\pi$ amplitude which allows us to continue it to the
second Riemann sheet. The amplitude contains (in spirit of the
linear $\sigma$-model) two bare objects: the resonance and the
background. The main idea is that a joint unitarization of two
objects should be described correctly by the field theory methods.
As concerning the form of a background contribution at the tree
level, it can be modelled by a maximally simple method.

There are different ways to construct such analytical amplitude.
We found the suitable one the formalism of the unitary mixing, the
obtained amplitude is analytical and unitary automatically. Such
construction is rather flexible which allows us to investigate
some different physical situations.

Note that from other side the bare pole
located at
$s <\ 0$ may be considered as some effective cross exchange and its value
$m_2^2 \sim -m_{\rho}^2$, obtained from a fit, confirms this interpretation.
As compared with standard N/D method our amplitude with two bare objects
automatically has a zero,
which is necessary to describe the S-wave low energy data.

\section{Formalism of the unitary mixing}

If there exist n bare states with the same quantum numbers then
the dressing of their propagators should account also the mutual
transitions between them. The process of joint dressing is
described in this case by the system of Dyson-Schwinger equations:
\begin{equation} \Pi_{ij}(s) =
\pi_{ij}(s) - \Pi_{ik}(s) J_{kl}(s) \pi_{lj}(s), \ \ \ \ \ \ \ \ \
i,j=1...n .
\label{Dyson}
\end{equation}

Here  $\pi_{ij}$ and $\Pi_{ij}$ are bare and dressed propagators
respectively, $J_{ij}$ are the self-energy contributions.

Let us consider mixing of two resonances (n=2) with one open
intermediate state. In this case the solutions of (\ref{Dyson})
are:
\begin{equation}
\Pi_{11}=\frac{D_{2}(s)}{D(s)}, \ \ \
\Pi_{12}=\frac{J_{12}(s)}{D(s)},\ \ \
\Pi_{22}=\frac{D_1(s)}{D(s)}.
\label{Pi}
\end{equation}
Here
\begin{eqnarray}
D(s)&=&D_1 D_2 - (J_{12}(s))^2,   \nonumber \\
D_1&=&m_1^2 - s - J_{11}(s),\ \ \ \ \ D_2=m_2^2 - s - J_{22}(s).
\label{}
\end{eqnarray}
In the case of scalar resonances interacting with pion pair, the
loops are of the form: \footnote{Note that we ignore the
subtraction constants in the loops. As it is shown in Appendix A a
subtraction polynomial in the loops can be removed by the
redefinition of bare parameters.}
\begin{equation}
J_{11}(s)=g_1^2 J(s),\ \ \   J_{22}(s)=g_2^2 J(s),\ \ \
J_{12}(s)=g_1 g_2 J(s).
\label{}
\end{equation}
\begin{equation}
J(s)=\frac{s-a}{\pi} \int_{4 m_{\pi}^2}^{\infty} \
\frac{ds'}{(s'-a)(s'-s)}\ \rho(s'),\ \ \ \ \  \
\rho(s)=\sqrt{\frac{s-4 m_{\pi}^2}{s}}, \label{J}
\end{equation}
where $0 < a < 4 m_{\pi}^2$ is the subtraction point, $g_i$ are
the coupling constants.

The $\pi\pi$ amplitude:
\begin{equation}
f=g_1^2 \Pi_{11}(s) + g_2^2 \Pi_{22}(s) + 2 g_1 g_2 \Pi_{12}(s) =
\frac{N(s)}{D(s)},
\label{amp}
\end{equation}
where
\begin{eqnarray}
D(s)&=&(m_1^2-s)(m_2^2-s) - J(s)N(s),  \nonumber  \\
N(s)&=&g_1^2(m_2^2 - s) + g_2^2(m_1^2 - s) .
\end{eqnarray}

It is evident that  (\ref{amp}) satisfies the elastic unitary
condition
\begin{equation}
Im\ f = \rho |f|^2 .
\end{equation}

The above equations can be applied not only for the case of two
resonances but also for the "resonance+background" situation, when
one of the bare poles is located at $s < 0$.  Just right this
situation arises for the S-wave I=0  $\pi\pi\to\pi\pi$ amplitude.
One can see from (\ref{amp1}) that our amplitude is zero at the
point:
\begin{equation}
s_0^0 = ( g_1^2 m_2^2  + g_2^2 m_1^2 ) / (g_1^2  + g_2^2),
\end{equation}
which should be $s_0^0 \sim m_{\pi}^2$ to reproduce the Adler
zero. So we have $m_2^2\ <\ 0$ \footnote{In spite of $m_2^2\ <\
0$, we keep using this notation to stress the presence of two
objects in the amplitude. Note by the way that our amplitude
(\ref{amp}) coincides except of notations with the amplitude of
Ref. \cite{VDS}, obtained from the low-energy bootstrap
equations.}.

Let  $m_2^2\ <\ 0$  be the bare zero of function  D(s), which
stays at the left of real axis after dressing. Then it is
convenient to subtract the loop at this point:
\begin{equation}
D(s)=(m_1^2-s)(m_2^2-s) - (J(s) - J(m_2^2)) \left [ g_1^2
(m_2^2-s)+ g_2^2 (m_1^2 - s) \right] .
\label{amp1}
\end{equation}
Below we shall use the amplitude (\ref{amp}), (\ref{amp1}) for
description of the experimental data. Here $m_1^2, g_1^2, m_2^2,
g_2^2$ are free parameters. The background contribution at the
tree level may be modelled by the pole or constant. It is
sufficient for successful description of the experimental data as
it is seen below.

\section{Analysis of $\pi\pi$ data in region of $m_{\pi\pi}<$ 900 MeV}

In the nearthreshold region we use the new data from $K_{l4}$
decay \cite{E865}, which may be seen at Fig.~\ref{Kl4} in
comparison with 1977 year data \cite{Ros}. We do not take into
account the old data \cite{Ros} as it has no practical effect on
the fit. The measured value in $K_{l4}$ decay is the phase shift
difference $\delta_0^0 - \delta_1^1$ thus we need an additional
information on the P-wave. We use for this purpose the
approximation of solution of the Roy equations from
Ref.\cite{Col01}. Fortunately, the $\delta_1^1$ contribution is
only about $1.5^{\circ}$ at the end of the interval due to the
P-wave threshold behavior, thus the uncertainty in $\delta_1^1$ is
negligible.
\begin{figure}[htb]
\begin{center}
\includegraphics*[height=6cm]{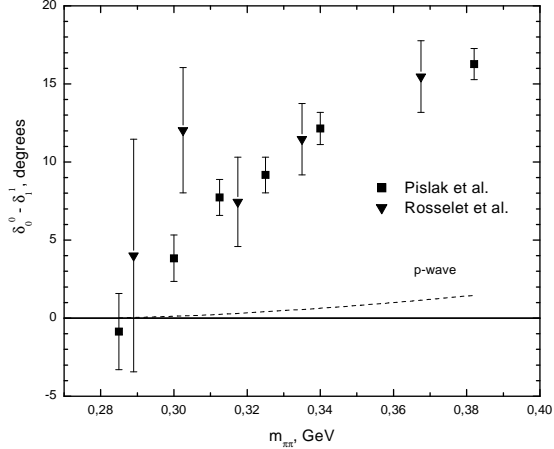}
\end{center}
\caption{Phase shift difference $\delta_0^0 - \delta_1^1$ from
experiment on $K_{l4}$ decay. }
\label{Kl4}
\end{figure}

Our main purpose is the $\sigma(600)$ resonance thus we restrict
ourselves by the energy region $m_{\pi\pi} < 0.9 $ GeV. It allows
us to use the one-channel approach and not to take into account
the $f_0(980)$ effect. In this region there exist different
experiments and different analyses of S-wave phase shifts, see
recent reviews \cite{Ver00,Ynd}.

Below we use only classical partial analyses of Protopopescu et
al. \cite{Pro} from $\pi^+p\to\pi^+\pi^-\Delta^{++}$ reaction and
Estabrooks and Martin one of CERN experiment \cite{Est74} on
$\pi^-p\to\pi^+\pi^-n$ (let us call their two solutions for
$\delta_0^0$ phase shift as EM I and EM II). Below we consider the
mentioned experimental data on the $\delta_0^0$ phase shift and
find very similar conclusions. As an example we focus in more
details on the the EM II solution.

Fig. \ref{EM2} displays the results of joint fitting of $K_{l4}$
data ($m_{\pi\pi} < 0.4 $ GeV) and EM II data ($0.51 < m_{\pi\pi}
< 0.9$ GeV). One can see that our amplitude (\ref{amp1}) describes
well these data.
\begin{figure}[htb]
\begin{center}
\includegraphics*[height=9cm]{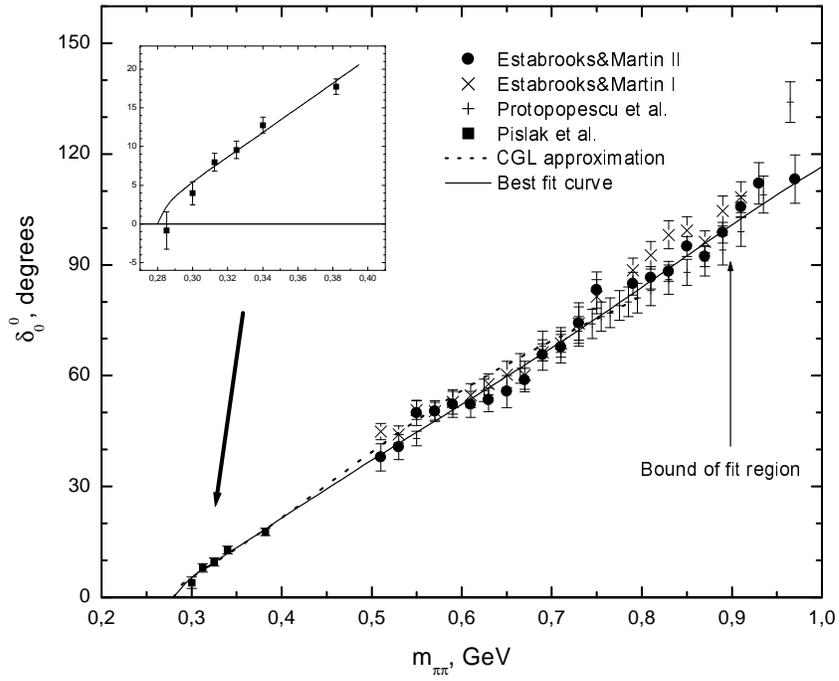}
\end{center}
\caption{Results of the fit of the $\pi\pi$ phase shift by the
amplitude (\ref{amp1}). Solid curve is the best fit to the data of
$K_{l4}$ decay and  EM II its parameters can be seen in the first
column of Table~\ref{summ}.} \label{EM2}
\end{figure}
\afterpage{\clearpage}

Best fit parameters are:
\begin{eqnarray}
m_1^2&=&0.659\pm0.041\ GeV^2 ,\  \  \  \  \ g_1^2=0.435\pm 0.036\ GeV^2,
 \nonumber  \\
m_2^2&=&-0.230\pm 0.114\ GeV^2 , \  \  \   g_2^2=0.177\pm 0.067\ GeV^2 ,
\nonumber  \\
\chi^2/DOF&=&17.7/21 .
\label{best}
\end{eqnarray}

Let us consider the zeros of function D(s) at the second Riemann
sheet\footnote{The values of the bare parameters have rather
limited meaning since they correspond to a given method of
renormalization. However the character of the pole movement is
more meaningful, at least when the loop contributions do not
dominate in amplitude.}. The procedure of analytical continuation
is described in Appendix B. The Fig. \ref{zeros} shows the zeros
location in the complex s plane corresponding to the best fit
parameters (\ref{best}). Let us stress that we find two zeros in
the $Re\ s > 0$ half-plane.
\begin{figure}[htb]
\begin{center}
\includegraphics*[height=6cm]{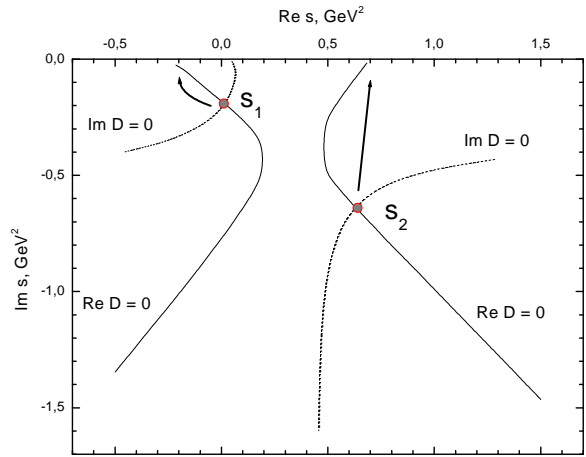}
\end{center}
\caption{Poles of the amplitude $\pi\pi\to\pi\pi$ with J=I=0 on
the second Riemann sheet at parameters (\ref{best}). Arrows
indicate the direction of the poles movement when the interaction
is turned off.}
\label{zeros}
\end{figure}
\afterpage{\clearpage} The Table \ref{summ} represents results of
the fit of the different low energy data by our amplitude
(\ref{amp1}). All data sets lead to the solutions with two poles:
close and distant \footnote{Our amplitude has a property
$f(s^*)=f^*(s)$ thus we have a pair of the complex conjugate poles
in the complex s plane. For definiteness we say about poles in the
$Im\ s > 0$ half-plane.}.
\begin{table}[ht]
\begin{center}
\begin{tabular}{|c|c|c|c|}     \hline

$K_{l4}$+EM II & $K_{l4}$+EM I & $K_{l4}$+Protopopescu
& CGL \cite{Col01} \\
$E<0.9$ GeV & $E<0.9$ GeV & $E<0.9$ GeV & $E<0.8$ GeV \\
                \hline
$m_1^2=0.659\pm0.041$ & $m_1^2=0.586\pm 0.025$ &
$m_1^2=0.794\pm 0.114$ & $m_1^2=0.845$ \\

$g_1^2=0.435\pm 0.036$ & $g_1^2=0.382\pm 0.020$
& $g_1^2=0.598\pm 0.151$ & $g_1^2=0.779$ \\

$m_2^2=-0.230\pm 0.114$ & $m_2^2=-0.113\pm 0.053$
& $m_2^2=-0.580\pm 0.405$ & $m_2^2=-0.573$   \\

$g_2^2=0.177\pm 0.067$ & $g_2^2=0.116\pm 0.030$
& $g_2^2=0.422\pm 0.331$ & $g_2^2=0.548$   \\

$\chi^2/DOF=17.7/21$ & $\chi^2/DOF=22.6/21$
& $\chi^2/DOF=5.3/19$ & $\chi^2/DOF=0$ \\   \hline

Poles: & Poles: & Poles: & Poles:  \\

$s_1=0.015+i\ 0.192$ & $s_1=0.045+i\ 0.132$
& $s_1=0.055+i\ 0.339$ & $s_1=0.104+i\ 0.250$   \\

$s_2=0.633+i\ 0.630$ & $s_2=0.632+i\ 0.533$
& $ s_2=0.484+i\ 1.020$ & $ s_2=0.659+i\ 1.620$   \\ \hline
\end{tabular}
\end{center}
\caption{Results of the fit of the different sets of experimental
data by our amplitude. $m_i^2$,  $g_i^2$, $s_i$ are in units of
$GeV^2$. In the last column our parameterization is compared to
approximation of the phase shift from Ref. \cite{Col01} (Solution
\cite{ACGL} of the Roy equations with use of scattering lengths
from the two-loop chiral perturbation theory calculations.). Our
phase shift practically coincides with  CGL approximation in this
energy region.} \label{summ}
\end{table}

In Fig.\ref{dif} we compare phase shifts corresponding to
different variants of Table  \ref{summ}.
\begin{figure}[htb]
\includegraphics*[width=0.47\textwidth]{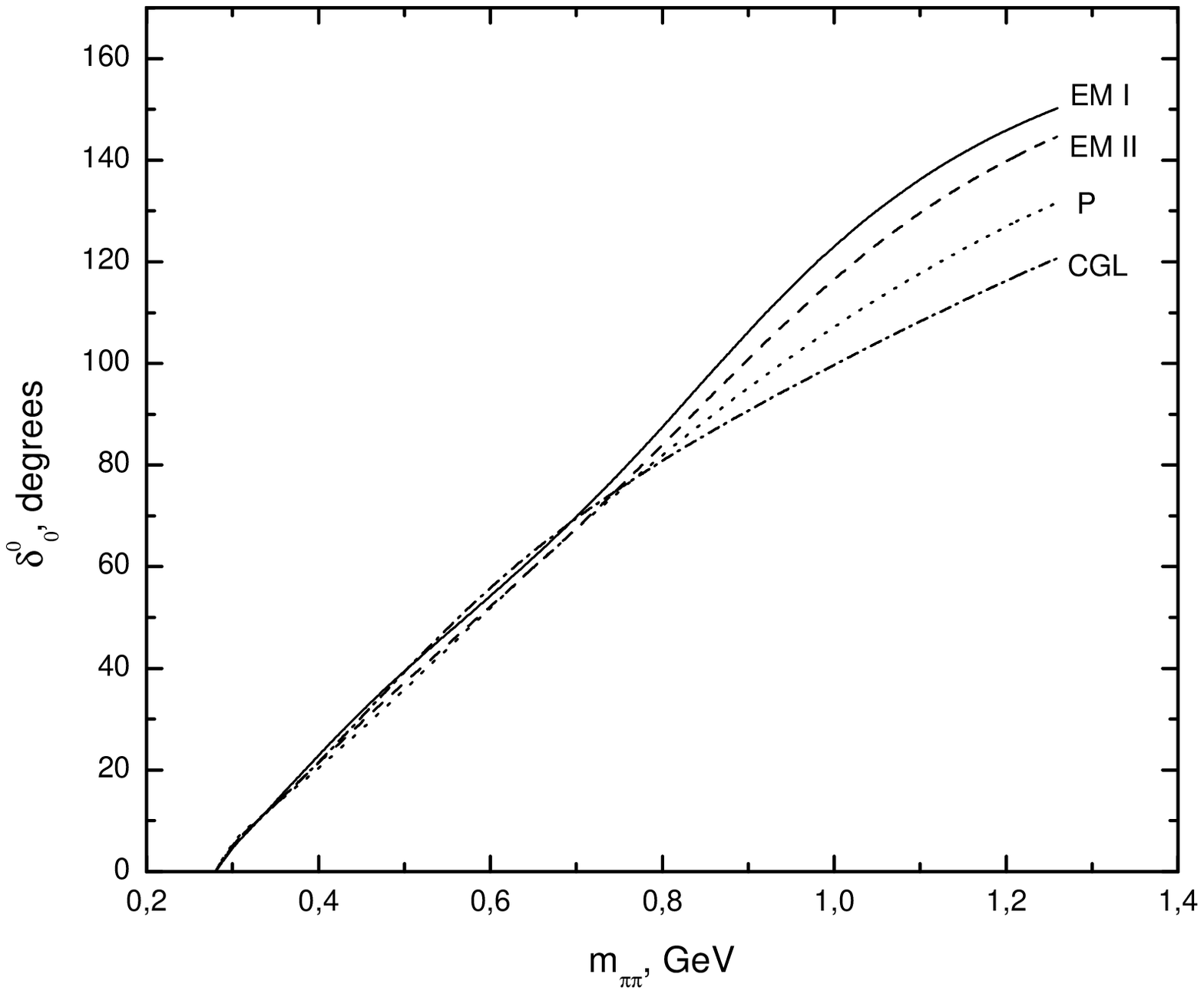}
\hfill
\includegraphics*[width=0.47\textwidth]{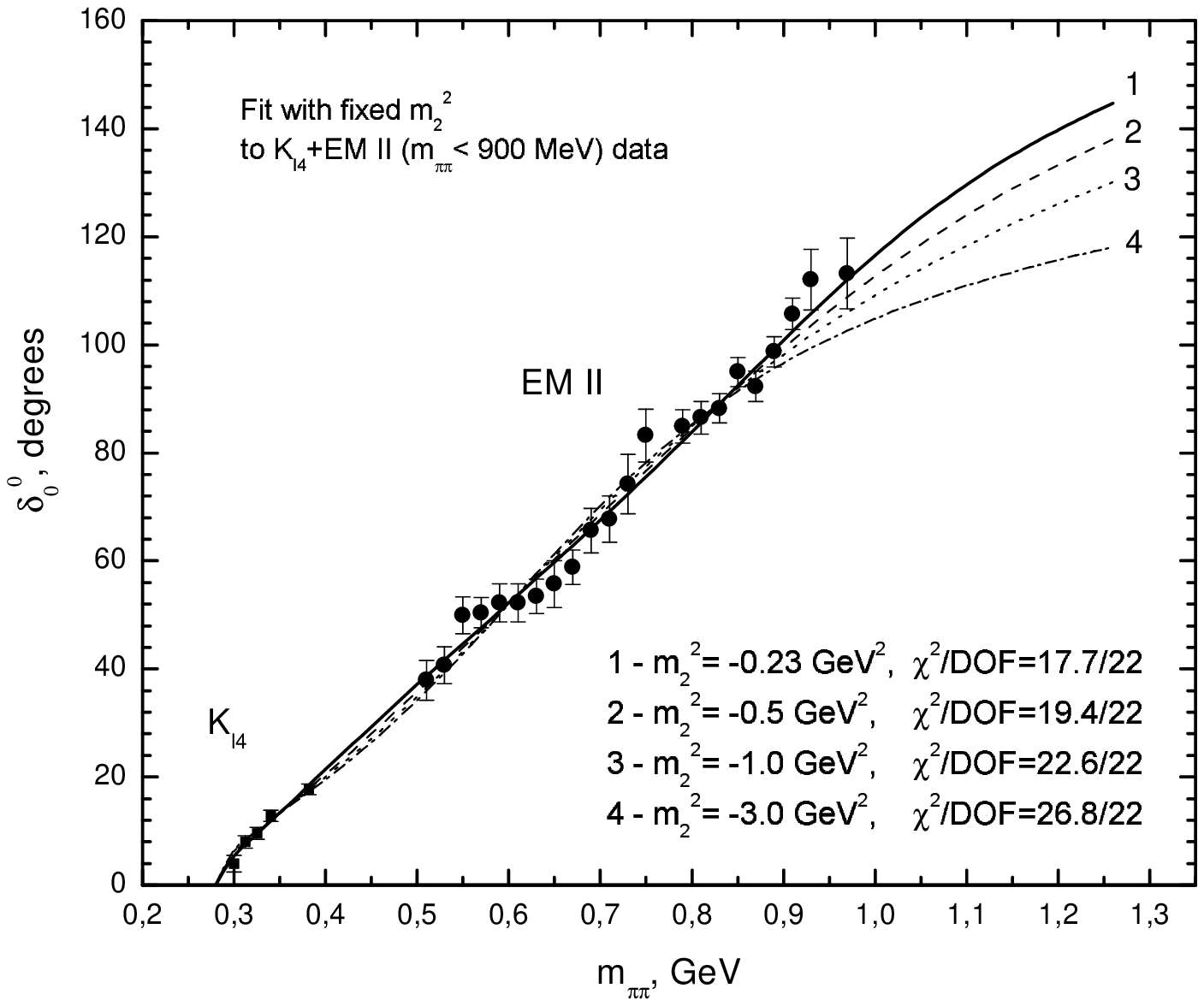}
\\
\parbox[t]{0.47\textwidth}{\caption{Phase shifts corresponding to different sets of
parameters shown in Table \ref{summ}.}\label{dif}}
\hfill
\parbox[t]{0.47\textwidth}{\caption{Data fit at fixed value of the left pole
$m_2^2$.} \label{fix}}
\end{figure}
%\afterpage{\clearpage}

We can see that our simple model (\ref{amp1}) corresponding to
joint unitarization of two bare objects: one pole at $s < 0$ and
another pole at $s > 0$ describes successfully the $\pi\pi$ phase
shift $\delta_0^0$ in the energy region below 900 MeV. We find two
poles in the complex s plane: one close to the origin with $Re\ s
\sim m_{\pi}^2$ and the second one with $Re\ s \sim 0.6\ GeV^2$.
However the behavior of the poles when interaction is turned off
$g_i^2\to 0$ is rather unexpected (see Fig. \ref{zeros}): the
close pole traditionally identified with $\sigma(600)$ meson moves
to the negative s region. While the second pole $s_2$ (most of
previous analyses did not observe it) tends to the real axis above
the threshold.

As an alternative we can investigate the case when the background
has not bare pole. It corresponds to the joint dressing in the
system "$\sigma$-pole + constant". For this purpose it is
sufficient to put the value $m_2^2$ negative and large in our
amplitude (\ref{amp1}).

In Fig. \ref{fix} there are shown the results of data fit with
different $m_2^2$ values. One can see that the experimental data
prefer rather close left pole $|m_2^2| < 0.6\ GeV^2$.

Fig. \ref{coup} illustrates the pole positions in the complex
plane at value $m_2^2=-1\ GeV^2$ and their behavior when
interaction is turned off. We observe that the behavior of poles
has been changed as compared with $m_2^2 > -0.5\ GeV^2$ case.
\begin{figure}[htb]
\includegraphics*[width=0.45\textwidth]{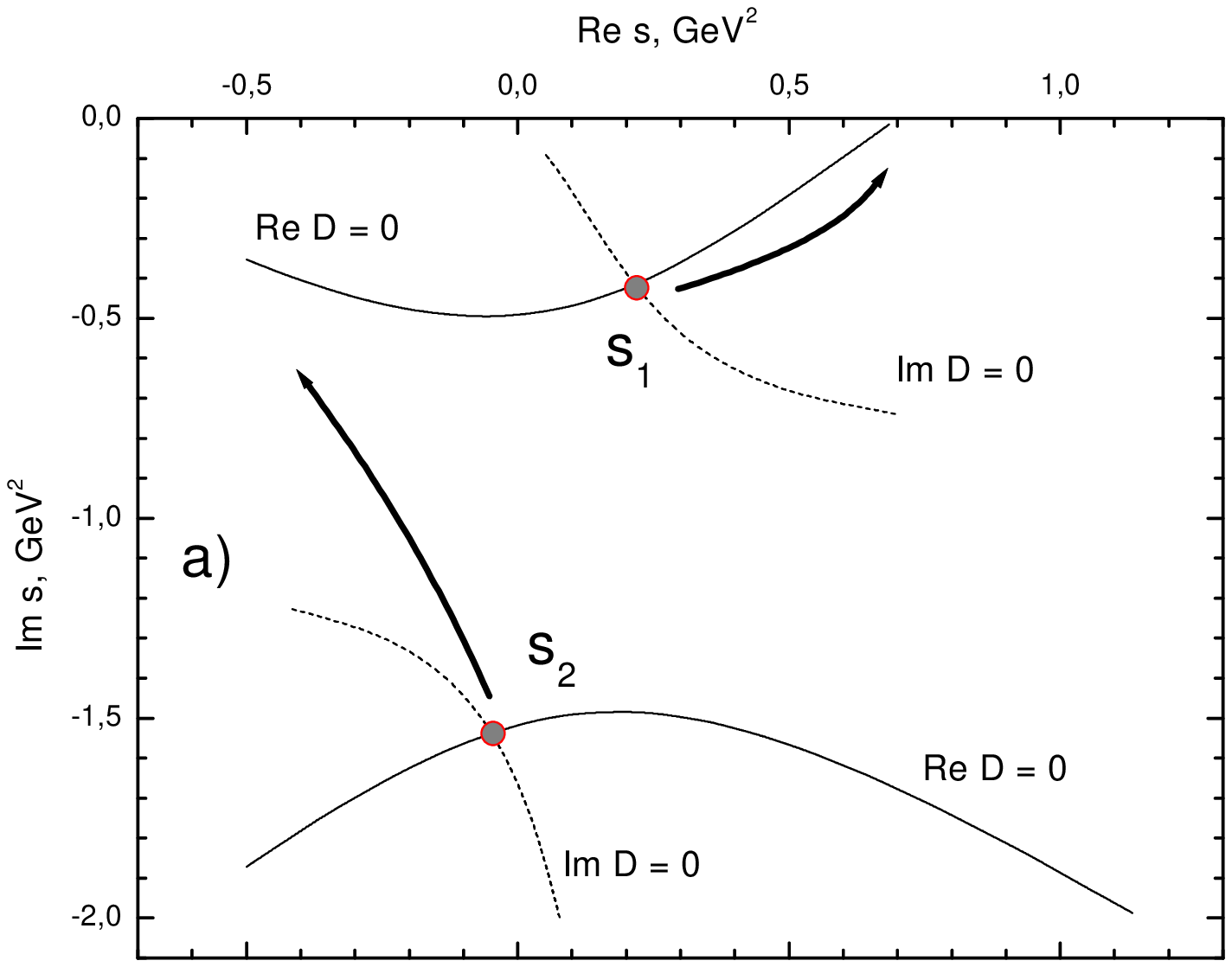}
\hfill
\includegraphics*[width=0.45\textwidth]{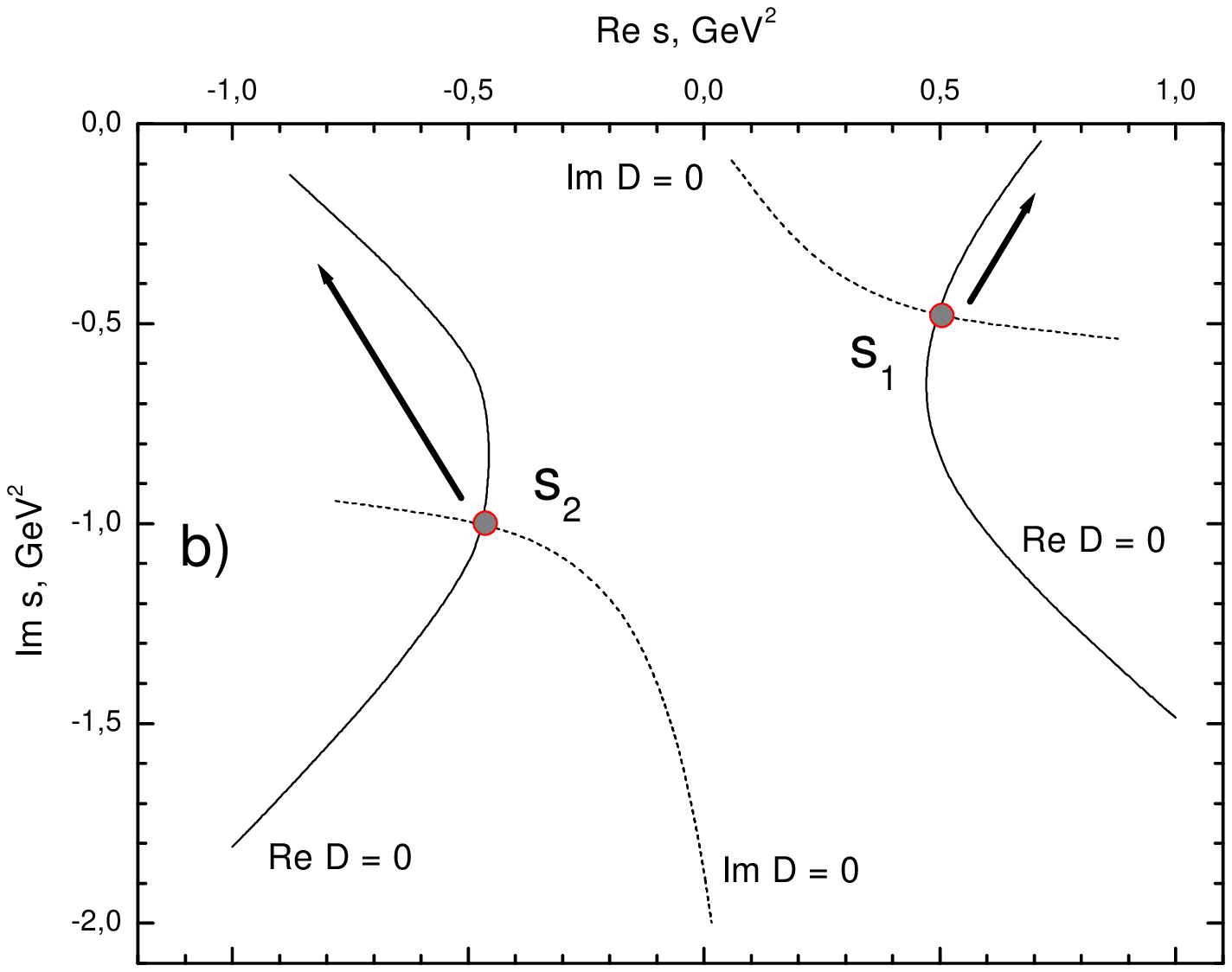}
\caption{ a)  Poles positions in the complex s plane at $m_2^2=-1\
GeV^2$ (fit of the data $K_{l4}$+EM~II) and their movement at $g_i\to 0$. \protect\\
b)  Illustration of the poles movement. As compared with a) we
slightly reduced the coupling constants $g_i^2\to 0.75 g_i^2$ with
fixed other parameters.}
\label{coup}
\end{figure}
%\clearpage \afterpage{\clearpage}

\section{Discussion}

We found that the $\pi\pi$ phase shift $\delta_0^0$ is well
described by a simple analytical amplitude (\ref{amp1}) in the
energy region from the threshold up to 900 MeV. Our amplitude
corresponds to a joint dressing of two bare objects: resonance and
background contributions. Background can be modelled either by a
pole with $Re\ s <0$ or by constant. As a next step one could
investigate the more complicated background model: left pole +
constant (just as in the linear $\sigma$-model). However since our
simple amplitude (\ref{amp1}) provides a good description of the
experimental data we suppose that inclusion of new degrees of
freedom has no meaning.

After the fit of the experimental data we found the presence of
two complex poles at the second Riemann sheet: one close to the
origin with $Re\ s_1 \sim m_{\pi}^2$ and the second one with $Re\
s_2 \sim 0.5 - 0.6\ GeV^2$. The close pole was seen in most of the
previous analyses of $\pi\pi$ scattering (its position is defined
mainly by Adler zero) and it was associated with the lightest
scalar meson $\sigma(600)$. Note that we approximated the
background term at the tree level by some pole, physically it
corresponds to the cross-exchange by $\rho$- or $\sigma$-meson.
The existing experimental data certainly prefer the pole form of
background as compared with constant.

As for behavior of the poles in the limit of $g_i \to 0$ we
observe that only the distant pole goes to the real axis at
positive s (see Fig.~\ref{zeros}). This fact holds true for all
variants indicated in Table~\ref{summ}. More detailed
investigation shows that such a behavior changes with $m_2^2$
value as it is schematically illustrated in Fig.~\ref{pole}. The
experimental data on $\pi\pi$ scattering with energy below 900 MeV
prefer the variant a) while the variant b) with $m_2^2 \sim -1\
GeV^2$ also can not be excluded (see Fig.~\ref{fix}). For example,
the found $\chi^2$ values are $\chi^2/DOF=17.7/22$ for a) and
$\chi^2/DOF=22.6/22$ for b) in one of variants of fit.
\begin{figure}[htb]
\includegraphics*[width=0.3\textwidth]{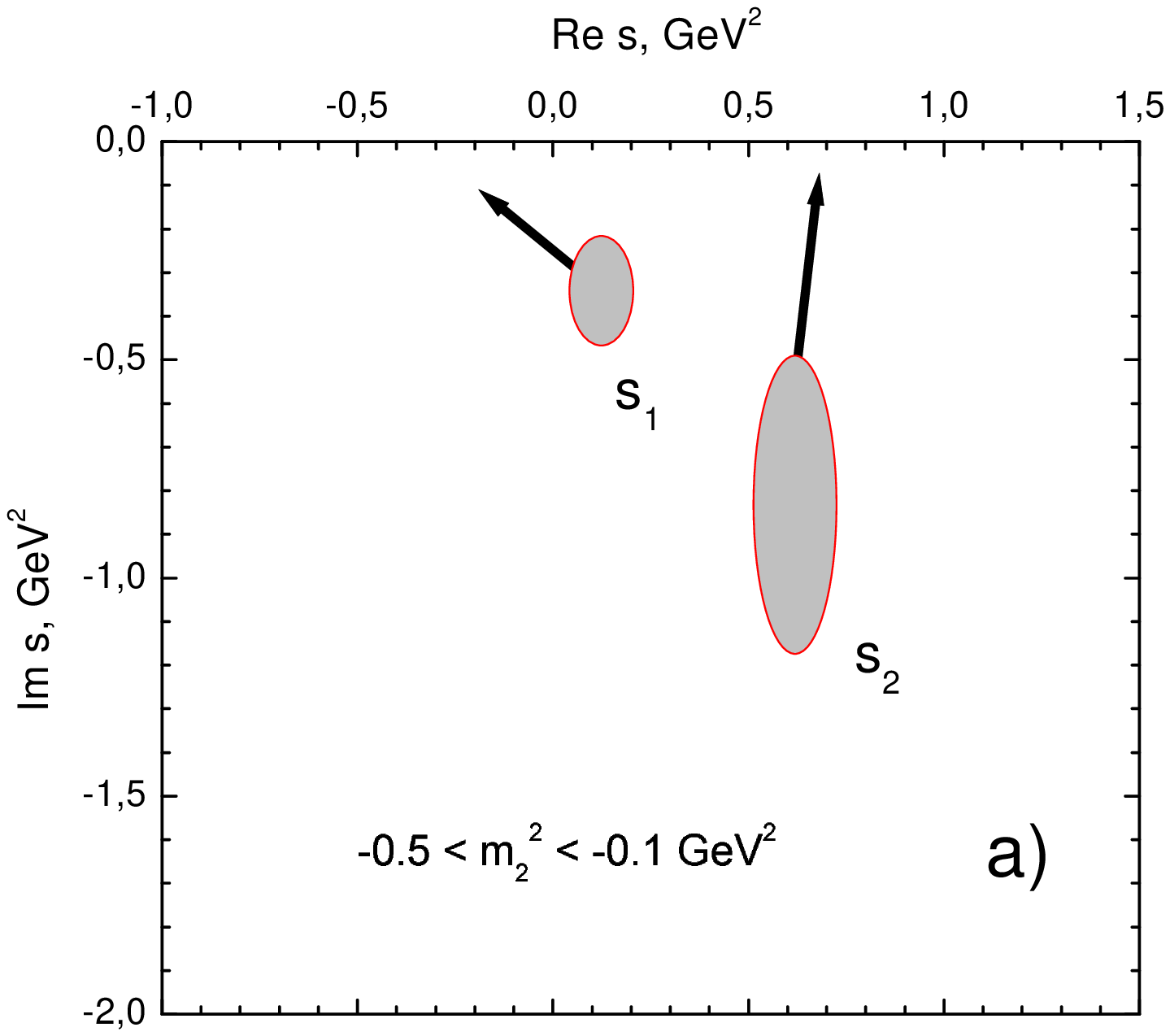}
\hfill
\includegraphics*[width=0.3\textwidth]{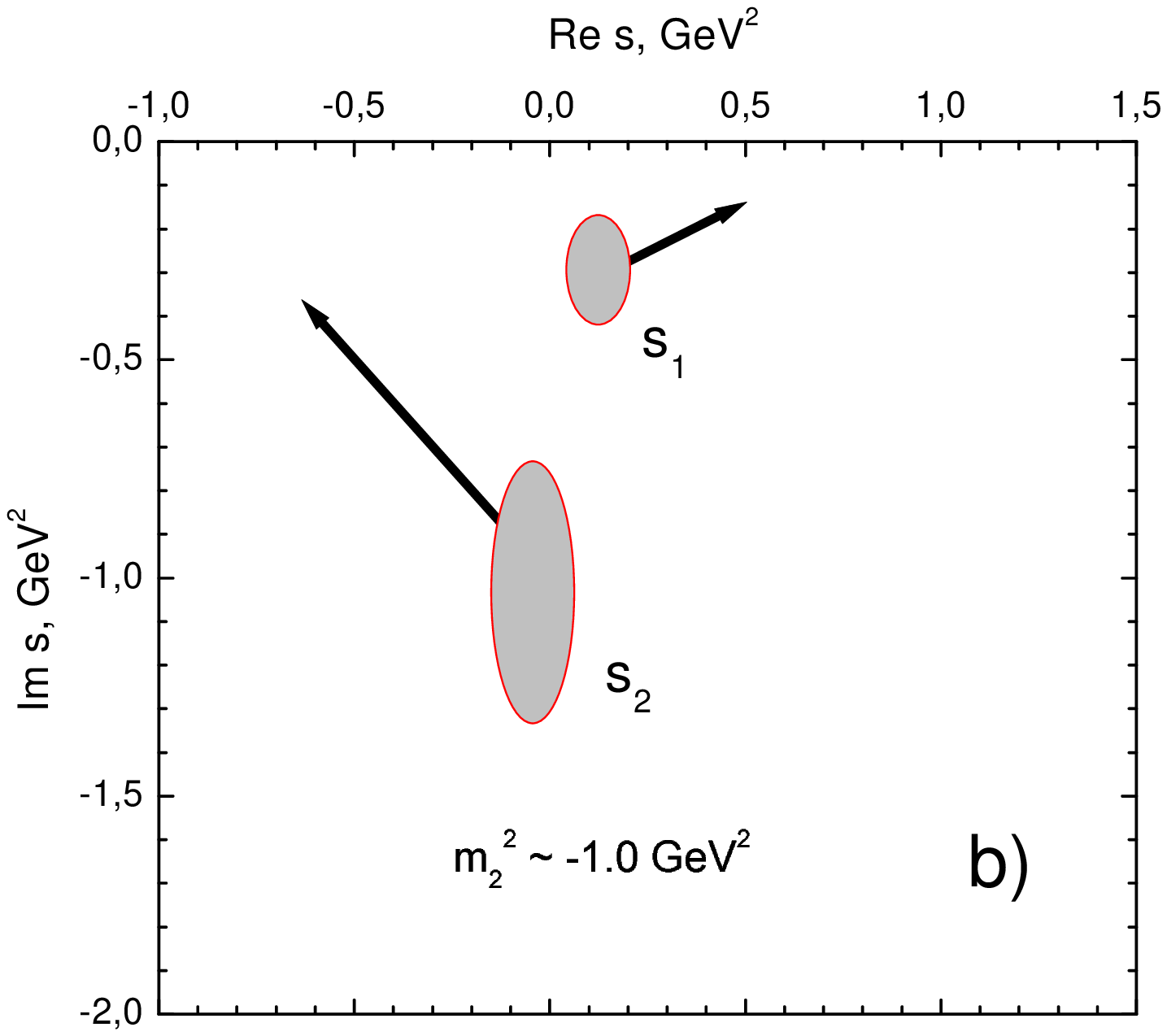}
\hfill
\includegraphics*[width=0.3\textwidth]{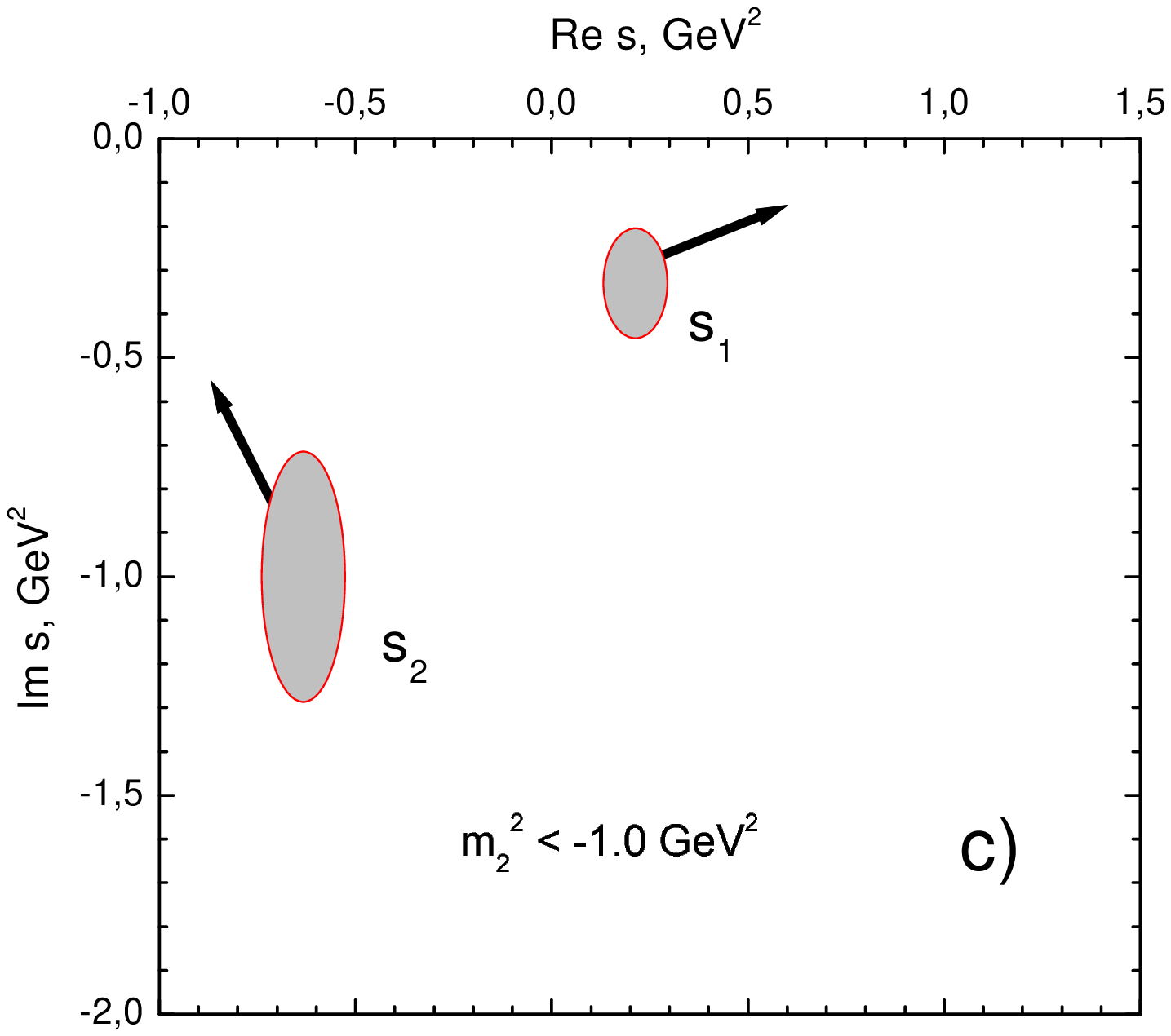}
\caption{Poles positions in the complex s plane and their movement
at $g_i\to 0$ is managed by $m_2^2$ value. The variant a) is
preferable by quality of the data description but b) also has an
acceptable value of $\chi^2$.} \label{pole}
\end{figure}

In view of discussion \cite{Tor96,Izg,Sha,Bog,Bev} whether the
$\sigma(600)$ the intrinsic state or it is dynamically generated,
our results should be interpreted as an indication for a dynamical
nature of the $\sigma(600)$. In this case the second pole should
be associated with intrinsic $q\bar{q}$ state having regard to
above remarks.

We suppose that the most interesting question is the meaning of
the second pole $s_2$. It was seen only in a few previous
analysis, e.g. in \cite{Ani00}, where it was considered as an
artefact since it was located out of the considered energy region.
In our analysis with account of the much more exact data from
$K_{l4}$ decay, this pole has moved to lower value $Re\ s_2 \sim
0.6\ GeV^2$. As for its imaginary part, it is rather uncertain
(see Table 1) and may be abnormally large for resonance state. We
suppose that further fate of this pole may be solved by an
analysis in the extended energy region.

In any case it is clear that in fact we have the joint complex
"$\sigma(600) + Background + f_0(980)$", which should be studied
jointly and by the adequate methods.

\newpage
\appendix
\section{Reparameterization of hadron amplitude}

Let us consider the unitary mixing of two bare poles with presence
of one intermediate state. We are interested in a number of
independent parameters in $\pi\pi\to\pi\pi$ amplitude. Let us
write it in a matrix form:
\begin{equation}
f=
\left(
\begin{array}{cc}
g_1 ,& g_2
\end{array}
\right)
\left(
\begin{array}{cc}
\Pi_{11},& \Pi_{12} \\
\Pi_{21},& \Pi_{22}
\end{array}
\right)
\left(
\begin{array}{c}
g_1 \\
g_2
\end{array}
\right)
\equiv { g^T \cdot \Pi \cdot g }
\end{equation}
Here $\Pi$  is the symmetrical matrix of propagator.

Let us start from the most general case when all loops have a
subtraction polynomial of a first degree \footnote{Higher degree
of polynomials leads to dominating of loops contributions at large
s. It leads to the changing of the problem's index and to changing
of number of poles as compared with the non-interactive case.}.
\begin{equation}
J_{ij}=g_i g_j (P_{ij}(s)+J(s))=g_i g_j (a_{ij}+b_{ij}s+J(s)),
\label{}
\end{equation}
where
\begin{equation}
J(s)=\frac{s-a}{\pi}\int_{4 m_{\pi}^2}^{\infty}
\frac{ds'}{(s'-a)(s'-s)}\sqrt{\frac{s'-4 m_{\pi}^2}{s'}}. \label{}
\end{equation}
Here a is subtraction point $0 < a < 4 m_{\pi}^2$, for analytical
continuation it is not convenient to subtract integral at zero.
There are ten parameters: bare masses $m_1, m_2$, coupling
constants $g_1, g_2$ and  6 subtraction parameters in the loops.

We can perform a transformation of propagators and coupling
constants, which does not change the amplitude:
\begin{eqnarray}
f&=&
 g^T\cdot S^{-1}S\cdot \Pi \cdot S^{-1}S\cdot g =
g^{'T} \cdot \Pi^{'} \cdot g^{'},  \nonumber  \\
\Pi^{'}&=&S\Pi S^{-1} , \ \ \ \ \ \ \ \ g^{'}=S g .
\label{transf}
\end{eqnarray}
Let us make few transformations consequently:
\begin{enumerate}
\item Firstly by the orthogonal transformation
\begin{equation}
S=\left(
\begin{array}{cc}
\cos{\theta}  & \sin{\theta} \\
-\sin{\theta} & \cos{\theta}
\end{array}
\right)  \nonumber
\label{}
\end{equation}
we delete the linear on s term in  the non-diagonal loop:
$b_{12}'=0$.
\item Then by the scale transformation
\begin{equation}
\Gamma=\left(
\begin{array}{cc}
\gamma_1  & 0 \\
0         & \gamma_2
\end{array}
\right)  \nonumber
\label{}
\end{equation}
we make the coefficient at s in $\Pi_{11}$, $\Pi_{22}$ by unity.
After it any orthogonal transformation can not generate again the
linear on s term in the non-diagonal loop.
\item We use one more orthogonal transformation to delete a subtraction constant
in the non-diagonal loop.
\item Finally, we can redefine the masses, absorbing the subtraction constants in
the diagonal loops.
\end{enumerate}

As a result we came to parametrization (\ref{Pi})-(\ref{J}) which
contains four parameters: masses and coupling constants.

\section{Analytical continuation of loop}

Let us consider the two-sheet analytical function:
\begin{equation}
F^{(n)}(s)=i\sqrt{\frac{s-4 m_{\pi}^2}{s}}=i \left| \frac{s-4
m_{\pi}^2}{s} \right|^{1/2}\ \frac{e^{i\ \varphi_1/2}}{e^{i\
\varphi_2/2}}\cdot (-1)^{(n-1)}, \ \ \ \ \ \ n=1,2.
\end{equation}
The cuts are chosen from $-\infty$ to zero and from $4 m_{\pi}^2$
to $+\infty$.

Let us write down the Coshi theorem on the first Riemann sheet
\begin{equation}
F^{(1)}(s)-F^{(1)}(a)=J^{(1)}+L(s).
\label{Coshi}
\end{equation}
$0<a<4 m_{\pi}^2$,
\begin{equation}
J^{(1)}=\frac{s-a}{\pi} \int_{4 m_{\pi}^2}^{\infty}
\frac{ds'}{(s'-a)(s'-s)} \sqrt{\frac{s'-4 m_{\pi}^2}{s'}}, \ \ \ \
L(s)=\frac{s-a}{\pi} \int_{-\infty}^0 \frac{ds'}{(s'-a)(s'-s)}
\sqrt{\frac{4 m_{\pi}^2-s'}{-s'}}.
\end{equation}
One can see from (\ref{Coshi}) that continuation of the loop to
the second Riemann sheet is performed as:
\begin{eqnarray}
J^{(2)} = -F^{(1)}(s) - F^{(1)}(a) - L(s) = \nonumber \\
= J^{(1)}-2F^{(1)} =  \nonumber \\
= -J^{(1)}-2(F(a)+L(s)).
\end{eqnarray}
The first expression seems the most convenient in numerical
calculations.

\end{rmfamily}
\end{document}